# PediDemi - A Pediatric Demyelinating Lesion Segmentation Dataset

Maria **Popa** [1], Gabriela Adriana Vișa [2]

**1** Department of Computer Science, Faculty of Mathematics and Computer Science, Babeș Bolyai University Cluj-Napoca, Romania, Mihail Kogălniceanu 1
**2** The Clinical Pediatric Hospital Sibiu, Pompeiu Onofreiu 2-4, Sibiu, Romania

**Abstract**
Demyelinating disorders of the central nervous system may have multiple causes, the most common are infections, autoimmune responses, genetic or vascular etiology. Demyelination lesions are characterized by areas were the myelin sheath of the nerve fibers are broken or destroyed. Among autoimmune disorders, Multiple Sclerosis (MS) is the most well-known Among these disorders, Multiple Sclerosis (MS) is the most well-known and aggressive form. Acute Disseminated Encephalomyelitis (ADEM) is another type of demyelinating disease, typically with a better prognosis. Magnetic Resonance Imaging (MRI) is widely used for diagnosing and monitoring disease progression by detecting lesions. While both adults and children can be affected, there is a significant lack of publicly available datasets for pediatric cases and demyelinating disorders beyond MS. This study introduces, for the first time, a publicly available pediatric dataset for demyelinating lesion segmentation. The dataset comprises MRI scans from 13 pediatric patients diagnosed with demyelinating disorders, including 3 with ADEM. In addition to lesion segmentation masks, the dataset includes extensive patient metadata, such as diagnosis, treatment, personal medical background, and laboratory results. To assess the quality of the dataset and demonstrate its relevance, we evaluate a state-of-the-art lesion segmentation model trained on an existing MS dataset. The results underscore the importance of diverse datasets for developing more robust models capable of handling a broader spectrum of demyelinating disorders beyond MS.

## 1. Background

Autoimmune demyelinating disorders affect the central nervous system and are determined by the presence demyelination lesions. These lesions are characterized by areas were the myelinsheath of the nerve fibers are broken or destroyed. Acute Disseminated Encephalomyelitis (ADEM) and Multiple Sclerosis (MS) are two common types of demyelinating disorders (Duignan and Hemingway, 2019).

ADEM is an acute inflammatory autoimmune disorder affecting the central nervous system and characterised by demyelination in the brain and spinal cord (Anilkumar et al., 2024; Xiu et al., 2021). ADEM affects mainly children and is among the most frequent demyelinating disorders in childhood (Xiu et al., 2021). Researchers found that one of the possible causes of ADEM could be the environmental stimulus, such as vaccination or infectious disease (Anilkumar et al., 2024; Cautilli et al., 2023).

Compared to Multiple Sclerosis, a chronic autoimmune disease, ADEM has a significantly better prognosis, with 57–89% of affected children achieving full recovery (Kumar et al., 2014). Its incidence is estimated at 0.07-0.51 per 100,000 children in Europe and 0.2-0.6 per 100,000 in North America (Xiu et al., 2021). Patients with MS more commonly develop lesions in the subcortical white matter, periventricular area, and corpus callosum, while patients with ADEM tend to have lesions in the cortical gray matter and deep gray matter, including the basal ganglia.

Diagnoses vary across patients, and MS and ADEM are not the only demyelinating disorders encountered in clinical practice. A variety of other demyelinating conditions exist, and in some cases, patients present with demyelinating lesions that do not clearly fit any one diagnostic category, placing them in a diagnostic gray zone.

Disorders such as Neuromyelitis Optica Spectrum Disorder (NMOSD) (Kim et al., 2012), Myelin Oligodendrocyte Glycoprotein Antibody-associated Disease (MOGAD), and infections (bacterial, viral, or parasitic) can also cause white matter abnormalities that mimic MS (Mukherjee et al., 2025). Moreover, lesion size, location, and progression can vary significantly across different demyelinating conditions, further complicating differential diagnosis and underscoring the importance of datasets that include a broader spectrum of demyelinating diseases.

Publicly available datasets for MS do exist and have supported significant advances in automated lesion detection. These include datasets released through major competitions such as MICCAI 2008, MSSEG 2016 (Commowick et al., 2021b), ISBI 2015 (Carass et al., 2017), MSSEG 2021 (Commowick et al., 2021a), and the upcoming MSLesSeg 2024 (Rondinella et al., 2025). Additionally, datasets like 3D-MR-MS (Lesjak et al., 2017) and a longitudinal MS dataset (Lesjak et al., 2016) have been made publicly available. However, datasets focused on non-MS demyelinating conditions remain scarce, and those representing patients with uncertain or evolving diagnoses are virtually non-existent.

In a previous study, we introduced PediMS (Popa et al., 2025b,a), a public pediatric MS dataset. PediMS includes longitudinal brain MRI scans of nine pediatric patients, each with 1–6 timepoints, all clinically diagnosed with MS. Given the rarity of pediatric MS compared to adult MS, datasets like PediMS play a valuable role in understanding early-onset disease.

In contrast, the dataset introduced in this study represents a broader and diagnostically heterogeneous population. It includes three patients with confirmed ADEM and several others who exhibit demyelinating lesions on brain MRI but do not meet diagnostic criteria for MS, ADEM, or other well-defined disorders. This new dataset captures the diagnostic uncertainty that often occurs in early or atypical demyelinating presentations, offering opportunities to study lesion characteristics, aid in differential diagnosis, and develop robust models for early detection.

This makes our proposed dataset complementary to PediMS. Rather than focusing on a specific diagnosis, it emphasizes radiological demyelination in ambiguous or evolving clinical contexts—an underrepresented but clinically relevant scenario in current neuroimaging resources.

This study introduces and publicly releases PediDemi (Popa and Visa, 2025), a pediatric demyelinating brain MRI dataset from The Clinical Pediatric Hospital, Sibiu, Romania. Unlike previous datasets, including our own PediMS, which focused exclusively on confirmed MS cases, PediDemi includes patients with demyelinating lesions who have not been diagnosed with MS. To our knowledge, it is the first public pediatric dataset to capture this diagnostically uncertain group. The dataset is intended to support the development of AI models that not only perform lesion segmentation in pediatric cases but also aid in distinguishing between MS and non-MS demyelinating presentations — a critical, underexplored challenge in clinical practice.

## 2. Summary

PediDemi contains brain MRI scans from 13 pediatric patients with non-MS demyelinating disorders, each with a single timepoint. For each patient, T1-weighted MPRAGE, T2, and FLAIR sequences are provided. Lesions were initially segmented by a junior rater and subsequently reviewed and corrected by a senior neuroradiologist to ensure consistency. A binary lesion mask aligned with the FLAIR modality is included for every subject.

This dataset is intended to advance lesion segmentation techniques and serve as a foundation for developing learning approaches that generalize beyond MS. By including an underrepresented population, PediDemi supports the development of more diverse and diagnostically inclusive classification models. To our knowledge, this is the first public pediatric dataset featuring patients with non-MS demyelinating disorders.

To demonstrate the dataset's utility and underscore the need for such data, we evaluated the performance of four models trained on public MS datasets and tested them on PediDemi. The dataset was preprocessed using the same pipeline described in (Badea and Popa, 2025). While some lesions were identified by the models, the results confirm that current models struggle to generalize to non-MS cases, highlighting the importance of dedicated datasets like PediDemi. The training code is based on the method proposed in (Badea and Popa, 2025) and is publicly available on GitHub: `https://github.com/marypopa/MS-Segmentation-Cross-Dataset`.

The remainder of the paper is organized as follows: Section 6 discusses the significance and limitations of the dataset; Section 3 details resource availability and potential applications; Section 4 outlines data collection and annotation procedures; Section 5 presents dataset validation and use-case testing; and Section 7 presents the conclusion and future work.

## 3. Resource availability

3.1 Potential Use Cases

The PediDemi dataset offers several potential applications for both clinical research and deep learning development:
**Lesion Segmentation in Non-MS Pediatric Cases:** As the first public dataset of pediatric patients with non-MS demyelinating disorders, PediDemi can be used to train and evaluate lesion segmentation models in a population that is underrepresented in current datasets.

**Cross-Dataset Generalization Studies:** The dataset enables researchers to assess how well existing MS-trained models generalize to atypical or non-MS demyelinating cases, revealing limitations in current approaches and promoting the development of more robust, clinically generalizable AI models.

**Diagnostic Support and Disease Differentiation:** Models trained using PediDemi may aid in distinguishing between MS and non-MS demyelination in pediatric populations, particularly in early or ambiguous cases where diagnosis is uncertain.

**Transfer Learning and Domain Adaptation:** Due to the overlap in imaging modalities with existing MS datasets, PediDemi can serve as a valuable resource for domain adaptation tasks, helping models trained on adult MS datasets adapt to pediatric non-MS contexts.

### 3.2 Licensing

The dataset is published under the CC BY license.

### 3.3 Data&Code Location

The dataset is stored in a Figshare repository and is available by accessing the following link: https://doi.org/10.6084/m9.figshare.28694435
Code and models can be found in the GitHub account at https://github.com/marypopa/, particularly in the repository: *pediDemi* and *MS-Segmentation-Cross-Dataset*.

### 3.4 Ethical Considerations

Data collection and use were approved by the hospital's Ethics Committee (approval no. ICR 3161/05.05.2025). Written informed consent for data sharing and future research use was obtained from the patients' parents or legal guardians.

## 4. Methods

Patient data and MRIs were first collected from the hospital's internal system. Each MRI scan underwent the same processing steps, as described in the following sections. After preprocessing, manual lesion segmentation was performed

### 4.1 Dataset collection

The *PediDemi* dataset includes 13 pediatric patients, each with a single brain MRI timepoint consisting of T1-weighted MPRAGE, T2-weighted, and FLAIR sequences. The dataset was constructed retrospectively from the imaging archive of The Clinical Pediatric Hospital Sibiu, covering cases from 2017 to 2022. The dataset includes patients aged between 3 and 17 years, comprising 8 males and 5 females.

All patients included in the PediDemi dataset met the following inclusion criteria: (i) age at symptom onset was 17 years or younger; (ii) demyelinating lesions were visible on brain MRI; (iii) no clinical diagnosis of Multiple Sclerosis was established at the time of imaging; and (iv) T1-weighted, T2-weighted, and FLAIR sequences were available from the same timepoint.

No exclusion criteria were applied. All cases that satisfied the inclusion criteria and had complete imaging data were included in the dataset.

The MRI scans for patient examinations were performed using a 3T Siemens Magnetom Skyra scanner and the characteristics can be seen in Table 1.

### 4.2 Patient data

The dataset includes a .CSV file containing extensive patient metadata, including gender, age at symptom onset, diagnosis, treatment, personal medical background, clinical evolution, and disease progression over time. Among the presented cases, three patients have fulfilled the criteria for ADEM, while the remaining patients represent a diverse spectrum of disorders, such as autoimmune encephalitis, postinfectious encephalitis, rhombencephalitis, migraine, and clinically isolated syndrome (see Table 2).

All patients, except those diagnosed with migraine, received corticosteroid treatment, with or without intravenous immunoglobulin (IVIG).

Personal medical history was included, as it can be a relevant factor in understanding disease context. For example, patient P1 had a history of prematurity, language delay, and bronchial asthma. In the case of P6, the patient's mother had a brain tumor; P7 had a growth hormone deficiency; and P10 had hydronephrosis.

To assess disease progression on MRI, we also include the MRI evaluation in time. While most patients had no visible lesions on follow-up scans, patient P3, showed stable demyelinating lesions in the spinal cord. This case also had 2 clinical episodes more than 3 months apart. There was a clinical suspicion of multiphasing ADEM or MS raised, but on the follow-up scans no new lesions were discovered, so diagnosed remained for now multiphasic encephalomyelitis.

Clinical evolution is documented in the extended metadata file. In general, patients showed good clinical outcomes with some exceptions. One of these patients presented with postinfectious meningo-encephalitis and acute symptomatic seizures developed epilepsy after five years. Another patient diagnosed with autoimmune encephalitis developed both demyelinating and axonal degeneration in the peripheral nerves. In this case, the elevated level of anti-GAD and VGKC antibodies is correlated with brain and peripheral nerve lesions.

Although this dataset includes only 13 cases, it captures a rare group of pediatric patients with non-MS demyelinating lesions. These cases are often underrepresented in public resources, and the dataset provides valuable ground truth for training models to differentiate early MS from other conditions like ADEM.

Table 1: Scanner characteristics, where TR is repetition time; TE - echo time, TI - inversion time, FA - flip angle

| Modality | TR(ms) | TE(ms) | TI(ms) | FA(°) | Thickness(mm) | Pixel resolution |
|---|---|---|---|---|---|---|
| T1-MPRAGE | 2000 | 2.27 | 900 | 8 | 1 | 0.98 x 0.98 x 1 |
| T2 | 8870 | 100 |  | 150 | 3 | 0.47 x 0.47 x 3.9 |
| FLAIR | 9000 | 81 | 2500 | 150 | 3 | 0.75 x 0.75 x 3.9 |

Table 2: Clinical data

| Patient | Gender | Age | Diagnosis |
|---|---|---|---|
| P1 | M | 5 | Postinfectious encephalitis with polyneuropathy |
| P2 | M | 3 | Postinfectious meningo-encephalitis. Acute symptomatic seizures |
| P3 | F | 16 | Multiphasic encephalomyelitis |
| P4 | F | 17 | Clinically isolated sndrome |
| P5 | M | 5 | ADEM |
| P6 | M | 13 | ADEM. Longitudinal extensive myelitis |
| P7 | M | 13 | Migraine |
| P8 | M | 3 | Autoimmune encephalitis. |
| P9 | F | 12 | Rhombencephalitis |
| P10 | M | 12 | Clinically isolated syndrome. Optic neuritis |
| P11 | F | 11 | ADEM |
| P12 | M | 5 | Encephalomyeloradiculitis. Ischemic stroke in right sylvian teritory. |
| P13 | F | 12 | Migraine |

### 4.3 Image processing

Each MRI underwent the same processing steps. Firstly, each patient scan was anonymized and defaced. Then, brain extraction was performed using a tool that works for all brain image modalities, SynthStrip (Hoopes et al., 2022; Kelley et al., 2024). The last step included fixing the field variations using N4-Bias Field correction from FSL (Smith et al., 2004).

### 4.4 Annotation process

The lesions were initially segmented by a junior rater and subsequently validated by a senior expert - pediatric neurologist. JIM 9.0[1] (Xinapse Systems Ltd., UK) was used for image annotation. This software is widely utilized across various domains for image labeling and analysis. The segmentation was performed on FLAIR sequences, with T1 and T2 images also examined to confirm lesion validity.

The labeling process can be seen in Figure 1 and can be divided into the following steps:

1. The processed FLAIR was loaded into the tool.

2. Load T1 and T2 in another tool, such as Biotronics 3D[2] and activate the fusion feature for a better view of lesions.

3. In JIM 9.0, from the Toolkits menu, select ROI Analysis and start performing the region of interest (ROI) labelling by selecting the Contour feature (see Figure 1 (1))

4. Correct and adjust the ROI

5. Move across slices, and select the lesions (see Figure 1 (2))

6. Save the selected ROIs in a file using save.

7. To produce the mask (see Figure 1 (3)):

   - Go in the JIM 9.0 menu and select from the Process tab, the Masker
   - Load the processed FLAIR image
   - Load the region of interest file
   - tick the Always create a binary image mask
   - Apply and save the image.

### 4.5 Data organisation

Each patient's data is stored in a separate subdirectory within the main dataset directory. Inside each patient's folder, there is a subfolder named **T1**(Timepoint 1), which contains both a **raw** data folder and a **preprocessed** folder, allowing researchers to apply their own processing steps if needed.

The **preprocessed** folder includes:

- Brain-extracted images, prefixed with *brain_*

- Skull-stripped segmentation masks, prefixed with *mask_*

- N4 bias field-corrected images, prefixed with *n4_*

- The lesion segmentation mask, named *Consensus*

---

1. https://xinapse.com/jim-9-software/

2. https://www.3dnetmedical.com/public/

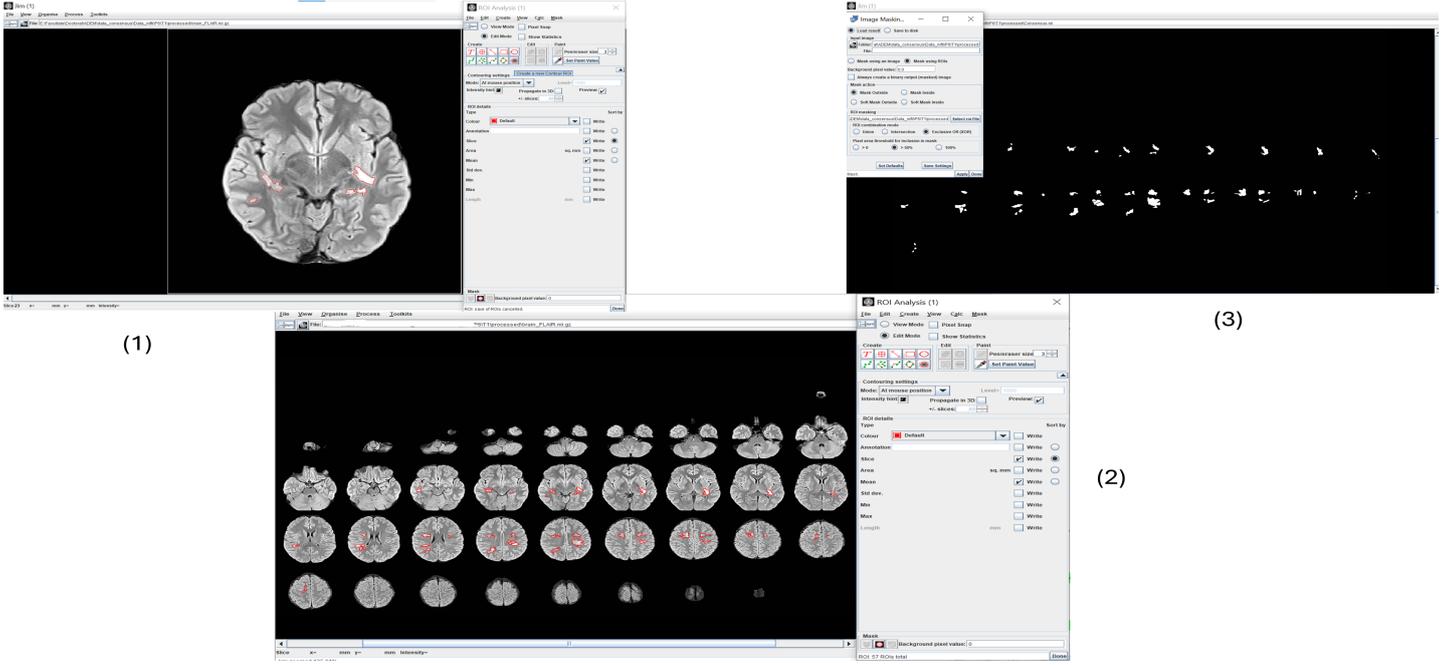

Figure 1: The annotation process: (1) presents a single image slice with the segmentation tool used for annotation (ROI Analysis); (2) displays all annotations across the entire MRI; (3) shows the creation of the binary mask using the Image Masking feature.

## 5. PediDemi validation

To ensure the high quality of the publicly released data, we utilized a state-of-the-art MS segmentation model developed in a previous study (Badea and Popa, 2025). This model is based on the U-Net++ architecture (Zhou et al., 2018). For further implementation details, refer to (Badea and Popa, 2025). We trained separate models on each public dataset and performed inference on the proposed dataset.

Table 3: Models run on the public datasets and tested on PediDemi

| Train on | Dice | IoU | F1 |
| --- | --- | --- | --- |
| MSSEG-2016 test | 0.34 | 0.22 | 0.39 |
| MSSEG-2016 train | 0.26 | 0.17 | 0.31 |
| 3D-MR-MS | 0.26 | 0.16 | 0.20 |
| ISBI-2015 | 0.20 | 0.12 | 0.28 |

### 5.1 Evaluation metrics

We used the Dice Similarity Coefficient (DSC), the Jaccard Index (IoU), and the lesion-wise F1 score (F1) in order to evaluate the models.

$$DSC = \frac{2TP}{2TP + FP + FN} \quad IoU = \frac{TP}{TP + FP + FN}$$

$$F1_{lesionwise} = \frac{2TP_{lesionwise}}{2TP_{lesionwise} + FP_{lesionwise} + FN_{lesionwise}}$$

where TP denotes True Positives, TN stands for True Negatives, FP represents False Positives, and FN corresponds to False Negatives.

### 5.2 Results

The models were trained independently on each public available dataset and tested on the PediDemi dataset. The numerical results can be seen in Table 3.

It can be observed that the model trained on the MSSEG-2016 dataset achieved the highest score. However, it remains far from a perfect score, which is necessary for clinical applicability. Among all models, this one detected the most lesions. A possible explanation is that MSSEG-2016 includes data from diverse imaging settings and patient populations. However, the model does not fully identify all lesions and misses some, particularly in ADEM cases, where lesion characteristics differ from those seen in MS, see Figure 5.2.

However, for patients who have not been diagnosed with ADEM, lesions are typically smaller and less frequent. As a result, the models introduced more false positives,

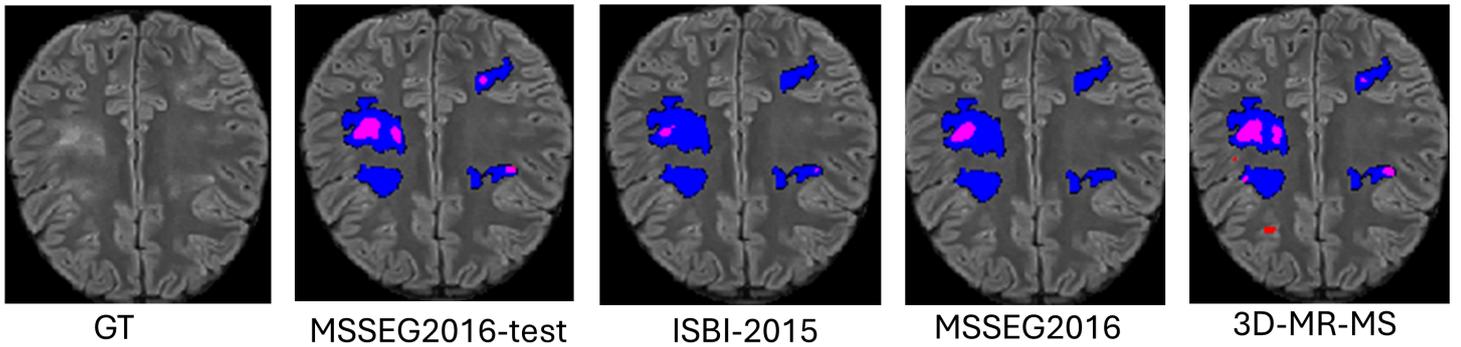

GT     MSSEG2016-test     ISBI-2015     MSSEG2016     3D-MR-MS

Figure 2: Segmentation results for Patient 6 (presented with ADEM). True Positives are shown in Magenta, representing regions correctly predicted by the model (i.e., where the prediction and ground truth overlap). Red indicates False Positives (regions predicted by the model but not present in the ground truth), and Blue indicates False Negatives (regions present in the ground truth but missed by the model). Images have been cropped for display purposes.

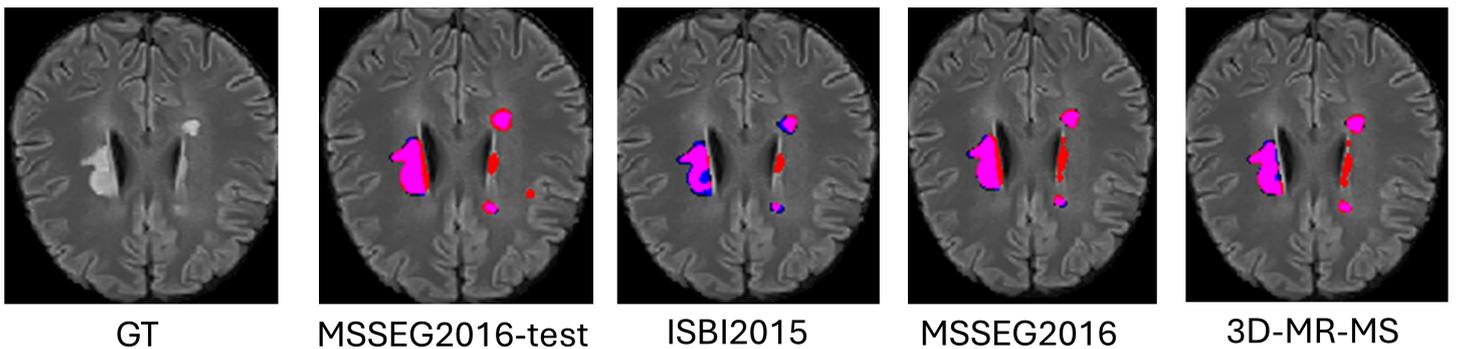

GT     MSSEG2016-test     ISBI2015     MSSEG2016     3D-MR-MS

Figure 3: Segmentation performed by the models for Patient 4. True Positives are shown in Magenta, representing regions correctly predicted by the model (i.e., where the prediction and ground truth overlap). Red indicates False Positives (regions predicted by the model but not present in the ground truth), and Blue indicates False Negatives (regions present in the ground truth but missed by the model). Images have been cropped for display purposes.

leading to lower overall scores. This is evident in cases such as Patient 4, see Figure 5.2.

## 6. Discussion

The PediDemi dataset fills a critical gap in the current landscape of neuroimaging resources by providing pediatric brain MRI scans of patients with non-MS demyelinating disorders-a population not represented in public datasets. Its availability can support the development of segmentation models that generalize beyond MS and improve differential diagnosis in pediatric neurology.

While the dataset is limited by its relatively small size and the absence of spinal cord or optic nerve imaging, it is valuable due to the rarity of the condition in pediatric populations.

PediDemi represents an important step toward more inclusive, real-world datasets that reflect the diversity of demyelinating presentations encountered in clinical practice.

## 7. Conclusion and future work

This paper presents PediDemi, a newly released pediatric dataset for demyelinating lesion segmentation. It contains MRI scans from 13 patients, including three with Acute Disseminated Encephalomyelitis (ADEM), and a diverse range of other non-MS demyelinating conditions. In addition to the imaging data, the dataset includes comprehensive patient metadata covering diagnosis, treatment, clinical history, and disease progression. To assess the dataset's quality and relevance, we evaluate the performance of models trained on existing public MS datasets when applied to PediDemi. The results underscore the importance of including more diverse patient populations and imaging contexts to improve the generalizability of lesion segmentation models. Future work will focus on developing models better adapted to pediatric clinical variability, as well as expanding the dataset to include spinal cord imaging to capture a more comprehensive view of demyelinating disorders.

## Ethical Standards

The work follows appropriate ethical standards in conducting research and writing the manuscript, following all applicable laws and regulations regarding treatment of animals or human subjects.

## Conflicts of Interest

We declare we don't have conflicts of interest.